\documentstyle[preprint,aps]{revtex}
\begin{document}
\draft
\title{Possible metal/insulator transition at $B=0$ in two dimensions}
\author{S.\ V.\ Kravchenko, G.\ V.\ Kravchenko\cite{gk}, and
J.\ E.\ Furneaux}
\address{Laboratory for Electronic Properties of Materials and
Department of Physics and Astronomy, University of Oklahoma,
Norman, OK 73019}
\author{V.\ M.\ Pudalov\cite{vp} and M.\ D'Iorio}
\address{National Research Council of Canada, IMS, Ottawa, Ontario,
Canada K1A OR6}
\date{October 15, 1993; resubmitted April 16, 1994}
\maketitle
\begin{abstract}
We have studied the zero magnetic field resistivity, $\rho$, of
unique high-mobility two-dimensional electron systems in silicon. At
very low electron density, $n_s$ (but higher than some
sample-dependent critical value, $n_{cr}\sim10^{11}\text{ cm}^{-2}$),
conventional weak localization is overpowered by a {\em sharp drop of
$\rho$ by an order of magnitude} with decreasing temperature below
$\sim1-2$~K. No further evidence for electron localization is seen
down to at least 20~mK. For $n_s<n_{cr}$, the sample is insulating.
The resistance is empirically found to scale with temperature both
below and above $n_{cr}$ with a single parameter which approaches
zero at $n_s=n_{cr}$ suggesting a metal/insulator phase transition.
\end{abstract}
\pacs{PACS numbers: 71.30.+h, 73.40.Qv, and 73.20.Fz}
\narrowtext

Although the problem of localization in disordered electron systems
has been studied both theoretically and experimentally for more than
a decade, there remain serious unanswered questions. Abrahams and
coworkers predicted \cite{aalr} that all the electron states in a
disordered 2D electron system (2DES) in zero magnetic field are
localized at zero temperature. This implies that there is no
metal/insulator (M/I) transition in an infinite 2D sample. Recently,
interest in this and related problems has intensified with studies of
the superconductor/insulator transition in ultrathin metal films
\cite{super,liu}. Furthermore, Azbel predicted \cite{azbel} that,
contrary to Ref.~\cite{aalr}, a system of noninteracting 2D electrons
in a model disorder potential with a random set of ``D-function''
scatterers at zero magnetic field and zero temperature is localized
{\em only} at energies below some mobility edge. At all energies
above this edge, extended states exist. According to Azbel
\cite{azbel}, the disagreement between his results and
Ref.~\cite{aalr} might indicate that the resistance strongly depends
on the range of the scattering centers. For the 2DES in silicon
metal-oxide-semiconductor field-effect transistors (MOSFET's),
Azbel's results might be applicable because the predominant
scatterers in these samples, particularly at low electron densities,
have been shown \cite{ando} to be short-range, similar to the model
potential.

In the beginning of 1980s, good agreement between theory \cite{aalr}
and experiment was achieved on samples with rather low mobility.
Experimental evidence for logarithmic decrease of conductance with
lowering $T$ was reported in Ref.~\cite{bishop,pepper}. For samples
with higher mobility, an approximately linear {\em increase} of
conductivity with decreasing temperature was found at temperatures
$\gtrsim1$~K \cite{cham,dolgopolov} at electron densities
$n_s\gtrsim4\times10^{11}$~cm$^{-2}$. This was explained by the
temperature dependence of the screening function for elastic
scattering \cite{stern,gold}. At low temperatures, $T\lesssim1$~K,
this increase in conductivity is again limited by weak localization
\cite{dolgopolov}.

Further improvement in the quality of samples has enabled access to a
qualitatively new level for this problem, where the electron-electron
interaction, rather than disorder, becomes the dominant parameter.
For instance, recent studies of the insulating behavior for new
ultra-high-mobility MOSFET samples \cite{prl} produced strong,
previously unobtainable evidence that the main mechanism for
localization at low $n_s$ in zero magnetic field is the formation of
a pinned electron solid due to these strong electron-electron
interactions.

Here we extend these studies to concentrate on slightly higher
electron densities where localization is absent at least down to
20~mK. We report interesting unprecedented behavior in similar
ultra-high-mobility (up to $7.1\times10^4$~cm$^2$/Vs) Si MOSFET's. At
zero magnetic field and at low electron densities (but higher than
some critical value, $n_{cr}\sim10^{11}\text{ cm}^{-2}$), we have
found that a conventional weak localization, observed at
$T\gtrsim1-2$~K, is overpowered by {\em a sharp drop of $\rho$ by an
order of magnitude} as the temperature is decreased. We then see no
signs of electron localization down to the lowest available
temperature, 20~mK. At $n_s<n_{cr}$, the resistivity monotonically
increases as $T\rightarrow0$, indicating an insulating state studied
extensively elsewhere \cite{prl}. At $n_s$ both below and above
$n_{cr}$ we have observed that the resistivity scales with
temperature with a single parameter.

Four samples from wafers with different mobilities have been studied:
Si$_{15}$ with maximum mobility, $\mu_{max}$, of
$7.1\times10^4$cm$^2$/Vs, Si$_{12}$ with $\mu_{max}=3.3\times10^4$
cm$^2$/Vs, Si$_{14}$ with $\mu_{max}=1.9\times10^4$ cm$^2$/Vs, and
Si$_{39}$
with $\mu_{max}=0.5\times10^4$ cm$^2$/Vs. Mobility as a function of
electron density for these samples is shown in Fig.~\ref{M(N)}. All
samples are rectangular with a source to drain length of 5~mm, a
width of 0.8~mm, and an intercontact distance of 1.25~mm. The
resistance was measured using a four-terminal dc technique with a
high input resistance DVM. For each sample we observed the same
$\rho(T)$ characteristics independent of contact configuration. The
$I-V$ characteristics of an electron gas are, in general, nonlinear
\cite{bishop,pepper,prl}. All data discussed here are within the
linear $I-V$ region.

Figure \ref{R(T)} shows $\rho$ versus temperature for Si$_{15}$, Si$_{12}$,
and Si$_{14}$ at different electron densities. At $T\gtrsim2$~K,
temperature dependencies of $\rho$ dependencies are rather weak:
$\rho$ increases slowly with decreasing temperature consistent with
weak localization [$\Delta \rho\propto\text{log}\; T$; see inset to
Fig.~\ref{R(T)}~(c)] for the four upper curves for each sample and
stays constant or decreases slightly for two lowest $\rho$ curves.
But as the temperature is further decreased, for all curves below
some ``critical'' $\rho(T)$ indicated by dotted lines in
Fig.~\ref{R(T)}, $\rho$ sharply drops {\em overpowering the onset of
localization visible at higher temperature}. Note that at
$T\lesssim1$~K, no further evidence for electron localization is seen
at temperatures down to 20 mK for the curves below the critical
lines. According to Ref.~\cite{pepper}, in low-mobility samples true
metallic behavior ({\em i.e.}, $\rho$ independent of temperature or
decreasing as $T$ decreases) never has been seen, and a weak increase
of the resistance with decreasing temperature is always present. In
contrast, for the curves below the critical lines, at least for Si$_{15}$
and Si$_{12}$, we observe {\em strongly} metallic behavior, a strong
decrease of $\rho$ with decreasing temperature. At the same time, for
the curves above the critical line, resistivity grows continuously
with decreasing temperature showing permanently localized state.

For all three samples shown in Figs.~\ref{R(T)}, one can see a
remarkable symmetry of $\rho(T)$ dependencies about the critical
lines, especially for the two curves adjacent to these lines. This is
reminiscent of flow lines around a repulsive fixed point at $T=0$
similar to that for the quantum Hall effect \cite{pru}. Similar
behavior has been also reported for the superconductor/insulator
transition in disordered metal films \cite{super,liu}. Note that the
critical lines for samples with different mobility tend to the same
$\rho\sim7\times10^4$~$\Omega$ as $T\rightarrow0$.

The low-$T$ behavior of $\rho(T)$ becomes less temperature dependent
with decreasing mobility: for example, for Si$_{14}$ the characteristic
relative drop of the resistivity is approximately 3 times weaker than
for Si$_{15}$. Eventually, for the lowest mobility sample, Si$_{39}$, the low
temperature drop does not exist (see inset in Fig.\ \ref{M(N)}). The
latter $\rho(T)$ is consistent with that observed in conventional Si
MOSFET's as reported in Ref.~\cite{pepper}.

We have performed a one-parameter scaling analysis for the
resistivity in the temperature region 350~mK to 4.2~K, above the
low-temperature saturation, for the curves lying both below
(``metallic'' side) and above (``insulating'' side) the critical
line. The results for the best sample, Si$_{15}$, are shown in
Fig.~\ref{scaling}. One can see that the resistivity can be written
in a scaled form, {\em i.e.}, $\rho(T,n_s)=\rho(T/T_0(n_s))$.
Resistivities for the metallic side collapse into a single curve
except for the curve closest to the boundary with
$n_s=0.89\times10^{11}$~cm$^{-2}$. Resistivities for the insulating
side similarly collapse into a single curve. The density dependence
of $T_0$ is shown in the inset. For both metallic and insulating
sides, $T_0$ falls sharply as $n_s$ approaches the critical electron
density, $n_{cr}\approx0.85\times10^{11}$~cm$^{-2}$. This scaling
analysis gives results similar to the beautiful results presented in
Ref.\ \cite{liu} for superconductor/insulator transition in
disordered Bi films.

The observed absence of localization for $n_s>n_{cr}$ at
$T\rightarrow0$, the sharp well-defined threshold between two types
of behavior, and the satisfactory single-parameter scaling of the
resistivity suggest a phase transition. This suggestion is consistent
with Azbel's theory \cite{azbel}. However, our results are not
conclusive evidence for the existence of a true M/I transition in a
2DES at zero temperature due to the finite sample size and finite
temperature.

It is impossible to explain the observed sharp drop of $\rho$ at low
$T$ with the same mechanism (temperature dependent screening)
suggested in Refs.\ \cite{dolgopolov,stern,gold} as the physical
cause for weak decrease in $\rho$ with decreasing temperature
observed at higher $n_s$ and $T$ \cite{cham,dolgopolov}. For
temperatures less than the collision broadening of the energy levels
\begin{equation}
T<T_c=\hbar/2\tau k_B\sim3\cdot10^4\text{
KVs/cm}^2\;\cdot\mu^{-1}
\end{equation}
the singularity in the dielectric function is washed-out by collision
broadening \cite{dassarma}, and  the temperature dependence of $\rho$
should disappear (here $\tau$ is the elastic scattering time). In our
situation, $\mu$ gets very low at low $n_s$ and therefore the cut-off
temperature, $T_c$, becomes very high, {\em e.g.}, $\sim30$~K for
$\mu=1\times10^3\text{ cm}^2/$V whereas the drop of $\rho$ is
observed at $T\lesssim1-2$~K.

Because the observed low temperature drop of $\rho$ is so large, and
because it overcomes weak localization, it is reasonable to assume
that it is caused by the destruction of the dominant scattering
mechanism for the 2DES. For low $n_s$, this mechanism is ionized
impurity scattering \cite{sg}. A typical density for ionized
impurities in high-mobility samples is $n_i\sim10^{10}\text{
cm}^{-2}$ \cite{dolgopolov,prl} which corresponds to an average
distance between charged scattering centers $\sim10^3\AA\gg
r_B\sim20\AA$, the Bohr radius. Therefore, these impurities can be
considered independent. Possible single-particle mechanism for the
destruction of the ionized impurity scattering could be as follows.
In principle, for $n_i\ll n_s$ one should expect a strong drop of the
resistance at temperatures below $T_b=E_b/k_B$ ($E_b$ is the binding
energy), where the charged scattering centers start to bind
electrons: in this case, the scatterers are neutralized by trapped
electrons, and, therefore, the scattering of residual free electrons
is much weaker than at $T>T_b$. The binding energy for a single
electron is $\sim40$~meV for Si MOSFET's, but screening by free
electrons strongly affects this figure. In Ref.~\cite{vinter}, where
the effect of screening was taken into account, the binding energy
was calculated to be a few tenths of meV, making $T_b\sim
T_{cr}\sim$~few~K conceivable but somewhat unlikely. In frame of this
model, it is also difficult to account for the scaling behavior of
resistivity and for the striking symmetry of $\rho(T)$ about the
dotted lines in Fig.~\ref{R(T)}.

Another physical cause for the observed drop in $\rho$ could be
electron-electron interactions which have the largest characteristic
energy at electron densities around $10^{11}\text{ cm}^{-2}$:
\begin{equation}
E_{\text{e-e}}\sim\frac{e^2}{\epsilon}n_s^{1/2}\sim5\text{ meV}\gg
E_F=\frac{\pi\hbar n_s}{2m^*}\sim0.6\text{ meV}\sim\hbar/\tau
\end{equation}
(here $E_{\text{e-e}}$ is the energy of electron-electron
interactions, $e$ is the electron charge, $\epsilon$ is the
dielectric constant, $E_F$ is Fermi energy, and $m^*$ is the
effective mass). In fact, there is a strong evidence \cite{prl} that
the insulating behavior at $n_s<n_{cr}$ is
caused by an electron solid formation due to these strong
electron-electron interactions. One could suppose that the state of
the system near the M/I transition, on the metallic side, is an
electron liquid dominated by a macroscopic multi-electron
wavefunction which suppresses scattering. The existence of such a
``liquid crystal'' was discussed earlier (see, {\em e.g.},
\cite{halp}) and recently has obtained strong experimental support
\cite{nicholas}. The symmetry of $\rho(T)$ depicted in
Fig.~\ref{R(T)} and common characteristic temperatures observed for
localized and extended state anomalies are indicators of a common
mechanism. This favors a many-body mechanism for the low temperature
resistivity drop. This effect would not exist in more disordered
samples such as Si$_{39}$ where disorder dominates the system at lower
densities destroying the coherence necessary to observe the
multielectron collective state.

Finally, we would like to note that the similar destruction of
already started localization by decreasing temperature was recently
observed at Landau level filling factor $\nu=1$ at very low $n_s$, in
the border of the existence of the quantum Hall effect \cite{nu1}.
There it was considered as evidence for the temperature-induced
sinking of the lowest extended state below the Fermi level as
$T\rightarrow0$; in this sense, the effect of temperature was
equivalent to the effect of the disorder. If one admits the existence
of the mobility edge in zero magnetic field (prohibited by the
scaling theory \cite{aalr} and predicted by Azbel \cite{azbel}),
similar ``sinking'' of the energy of the mobility edge with
decreasing temperature can cause the dramatic drop of $\rho$ reported
here.

We acknowledge useful discussions with B.~Mason, K.~Mullen,
D.~Khmelnitskii, and A.~Shashkin and experimental assistance from A.~
Japikse, J.~Dotter, and D.~Brown. We are especially thankful to
Lizeng Zhang who suggested to carry out the single-parameter scaling
analysis. One of us (S.V.K.) would also like to acknowledge
stimulating discussions with S.\ Semenchinsky. This work was
supported by grants DMR 89-22222 and Oklahoma EPSCoR via LEPM from
the National Science Foundation, grant 93-0214235 from the Russian
Fundamental Science Foundation, a Natural Sciences and Engineering
Research Council of Canada (NSERC) operating grant, and a grant from
the Netherlands Organization for Science, NWO.

\begin{figure}
\caption{Mobility vs $n_s$ for different samples at $T=20$~mK (Si$_{15}$
and Si$_{39}$) and 60~mK (Si$_{12}$ and Si$_{14}$). Inset shows $\rho(T)$ for
Si$_{39}$ at several $n_s$.}
\label{M(N)}
\end{figure}
\begin{figure}
\caption{Resistivity vs $T$ for electron densities near $n_{cr}$ for
Si$_{15}$ (a), Si$_{12}$ (b), and Si$_{14}$ (c). Inset shows a temperature
dependence of $\rho$ consistent with weak localization
($\Delta\rho\propto \text{log}\; T$) at temperatures above the drop
of $\rho$.}
\label{R(T)}
\end{figure}
\begin{figure}
\caption{Scaling behavior of the resistivity for Si$_{15}$. Inset shows
density dependence of the scaling parameter, $T_0$.}
\label{scaling}
\end{figure}
\end{document}